\RequirePackage{fix-cm}
\documentclass[twocolumn]{svjour3}          
\smartqed  

\usepackage{cite}
\usepackage{amsmath,amssymb,amsfonts}
\usepackage{algorithmic}
\usepackage{graphicx}
\usepackage{textcomp}
\usepackage{xcolor}
\usepackage{graphicx}
\usepackage{subcaption}
\usepackage{csquotes}
\begin{document}
	
	\title{Optical Networking in Future-land: From Optical-bypass-enabled to Optical-processing-enabled Paradigm  
	}
	
	
	\author{Dao Thanh Hai   
	}
	
	
	\institute{F. Author is working as an independent researcher \\
		\email{haidt102@gmail.com} 
	}
	
	\date{Received: date / Accepted: date}

	\maketitle
	
	\begin{abstract}
		The evolution of optical networks is enabled by both technological and architectural advances with the goal of reducing operational and capital expenditure per transmitted bit. While the former one stimulates significant system capacity, the latter paves the way for reducing the effective traffic load in network so that more traffic can be carried. Accordingly, optical node architectures have been transitioning from optical-electrical-optical mode to all-optical one, leveraging the scalability and efficiency of fully optical cross-connecting. Conventional wisdom in designing and architecting such switching nodes is nevertheless rooted in the intuition that when an optical channel crossing an intermediate node, it should be maximally isolated from other optical channels in order to avoid interference which may result in degrading signal quality. Such long-established paradigm perceiving the interference of optical channels transiting at the same node as an adversarial factor and should therefore circumvent, albeit reasonable, may leave vast unexplored opportunities. Indeed, the rapid advances in all-optical signal processing technologies has brought opportunities to re-define the optical node architecture by upgrading its naive functionalities from simply add/drop and cross-connecting to proactively mixing optical channels in photonic domain. Specifically, all-optical channel aggregation and de-aggregation technologies have been remarkably advancing in recent years, permitting two or more optical channels at lower bit-rate and/or modulation formats could be all-optically aggregated to a single channel of higher-rate and/or higher-order modulation format and vice versa. Such evolutionary technique is poised to disrupt the existing ecosystem for optical network design and planning, and thus necessitates for a radical change to unlock new potentials. To that end, this paper presents a new paradigm for future optical networks, namely, \textit{optical-processing-enabled networks}, which are powered by in-network all-optical mixing capability. We introduce the operational principle of optical channel aggregation (de-aggregation) and show how spectrally beneficial such innovative operations could yield by an illustrative example. Next, in order to maximize the aggregation opportunity, we present a mathematical model for optimal routing based on integer linear programming model. Numerical results on the realistic network topology COST239 are provided to quantify the spectral gain of aggregation-aware routing compared to the conventional one.  
		
	\end{abstract}
	
	\section{Introduction}
	\label{intro}
	
	Broadband Internet has become necessity for all people globally and most of us might not imagine how we would work, interact with others and shopping without it. However, less known is that the majority of Internet traffic have been carried out in optical fiber and indeed, optical communication and networking have become the foundation of global digital connectivity. In handling the continued explosive Internet traffic growth that doubles roughly every two years \cite{Cisco20}, numerous revolutionary technological and architectural solutions have been proposed and implemented to increase the system capacity on one hand, and on the other hand, to decrease the capital and operation costs. Major breakthroughs include coherent transmission, higher-order modulation formats, spectrally and spatially flexible optical transmission and recently wide-band optical transmission \cite{20years, ir4}. For many years, the capacity of an optical fiber has been considered as being nearly infinite and indeed, the system capacity of a single fiber in optical core networks has undergone leap-and-bound growth with a spectacular rise from the milestone of 1 Tb/s achieved in around 1996 to 319 Terabit/s across 3000 km recently set by NICT in 2021 \cite{NICT}. In a quarter of century, the system capacity has attained more than 300-fold growth and such achievements have been owing to the convergence of remarkable advances in electronic, photonic and digital signal processing technologies. However, enlarging the transmission capacity is just one side of operator's solutions to accommodate unprecedented traffic increases. The other side which is at least equally important to operators is on a desire to exploit such continuously expanding capacity in a cost and energy-efficient manner through innovative approaches in network architectures. \\ 
	
	From architectural perspectives, optical networking has undergone a paradigm shift in the 2000s time frame with a transition from optical-electrical-optical mode to optical-bypass and/or all-optical operations \cite{Simmons}. Such transition has been first driven by the observation that the majority of the traffic entering an optical node are transitional and therefore, can be kept in optical domain as they traverse the intermediate node rather than be electronically processed. Second the significant advances in optical switching technologies coupled with long-haul transmission has been instrumental to turn the observation into reality. Indeed, the arrival of optical-bypass-enabled networks have permitted the elimination of massive optical-electrical-optical interfaces and thus, resulting in substantial savings in both capital and operational expenses. As such, optical-bypass networking has greatly matured since the early 2000s and is now a well-adopted technology that have been deployed widely by carriers in both metro and backbone networks \cite{all-optical}. In optical-bypass networks, it is noted that the key functionalities of an optical node are simply to add/drop traffic and optically cross-connect transitional lightpaths. Most importantly, the foundation of optical-bypass networking paradigm is relied on the conventional wisdom that when an optical channel crossing an intermediate node, it should be maximally isolated from other optical channels in order to avoid interference which may result in degrading signal quality. Such long-established paradigm perceiving the interference of optical channels transiting at the same node as an adversarial factor and should therefore circumvent, albeit reasonable, may leave vast unexplored opportunities. In fact, the rapid advances in all-optical signal processing technologies has brought up unique opportunities to re-define the optical node architecture by upgrading its naive functionalities from simply add/drop and cross-connecting to pro-actively combining transitional optical channels in photonic domain so as to better utilize spectrum capacity \cite{optical_processing_1, optical_processing_2}. \\
	
	The capability to process signal directly in optical domain rather than converting it to electrical has been attracting the research communities for many years thanks to its super-efficiency in cost and energy and also its scalability to higher bit-rates \cite{optical_processing_1, optical_processing_2, optical_processing_3}. Although fiber nonlinearity has been usually referred as a major transmission impairment in optical fiber communications \cite{suggested1, suggested2}, its other uses have been extensively exploited for realizing various optical signal processing techniques. Among optical signal processing functions, all-optical aggregation of lower-capacity channels into a single higher-capacity channel and de-aggregation of a higher capacity channel into many lower-capacity channels have been advancing significantly in recent years from both theoretical, experiments and proof-of-concept demonstrations \cite{aggregation1, aggregation2, ref1_22, ref2_22, ref3_22}. The arrival of such functionality performing fully in optical domain paves the new way for improving the re-configurability and flexibility of optical networks in an attempt to attain greater spectral efficiency. Specifically, incorporating all-optical channel aggregation and de-aggregation at intermediate optical nodes re-defines the networking architecture as it challenges the long-established foundation of avoiding unwanted interference between transiting optical channels. Rather than simply performing add/drop function and/or cross-connecting individual optical channels as in optical-bypass networking, optical channels transiting the same intermediate node have new opportunities of being added together in photonic domain for greater spectral efficiency. \\
	
	In this paper, we propose a future-forwarding architectural paradigm, entitled, \textit{optical-processing-enabled} for optical networks by leveraging the in-network all-optical mixing capability. In Sect. 2, related works and contributions are highlighted. We then introduce the operational principle of optical channel aggregation de-aggregation in Sect. 3. How spectrally beneficial such innovative operations could yield is featured by an illustrative example in Sect. 4. Next, in order to maximize the aggregation opportunity, we present a mathematical model for optimal routing based on integer linear programming model in Sect. 5. Section 6 is dedicated for numerical results on the realistic network topology COST239 quantifying the spectral gain of aggregation-aware routing compared to the conventional one. Finally, the conclusions and perspectives are discussed in Sect. 7. \\ 
	
	\section{Related Works and Our Contributions}
	Optical-bypass networking have been well-developed in the last two decades with several research works for sustaining explosive traffic growth \cite{all-optical}. A central problem to solve in optical-bypass-enabled networks is the routing and resources allocation \cite{Algorithm1} and algorithms for solving such problem have been comprehensively investigated \cite{Algorithm2, Algorithm3, hai_iet} addressing various objectives such as spectrum efficiency, throughput, congestion level and energy consumption. The elastic optical technologies with optical-bypass architecture have been proposed and studied  \cite{EON3, EON4} as a substitution for fixed-grid optical networks toward greater spectrum efficiency. Various optical layer protection techniques including dedicated and shared protection have been developed and implemented by relying on optical-bypass architecture \cite{hai_csndsp, hai_ps2, hai_wiley}. Recently, space division multiplexing (SDM) technologies such as multicore fiber (MCF) and multi-mode fiber (MMF) have been intensively researched and in the context of optical-bypass framework, various algorithms for solving routing, spectrum, and core and/or mode assignment problem have been proposed to optimally exploit the expanded capacity enabled by SDM \cite{sdm1, sdm2}. In an effort for enhancing optical-bypass-enabled networks, proposals have been made by incorporating simple optical signal processing including wavelength/spectrum conversion, regeneration, and format conversion and these changes have been helpful in boosting the network performance \cite{regenerator1, regenerator2, regenerator3}. However, the performance improvement has come at a cost of more complicated network design problems. Moreover, it should be noticed that such adjustment has been limited to the optical manipulation of single individual transiting lightpath and therefore, missed the potential benefits derived from a wide range of signal processing enabled by controlled interference between through lightpaths. \\
	
	In an attempt to introduce various signal processing functions for optical channels crossing an intermediate nodes, network coding \cite{NC} (NC) has been  used in optical networks to attain greater throughput, security and capacity. The underlying idea is that intermediate nodes, instead of simply storing and broadcasting data as in conventional networking paradigm, is allowed to manipulate data and then forward such (non-) linearly mixed data to its output. Different from broadcast nature of signals in wireless networks, mining the bandwidth benefits of NC in optical networks have nevertheless required multiple signals that can be favorably mixed together and such condition often emerges from multi-cast scenarios \cite{nc-general3, nc-ofdm}, passive optical networks \cite{NC-PON}, or the case of disjoint paths in protection \cite{Kamal} which are prime targets of network coding research in optical networks. The majority of these works were however restricted to electronic network coding in combination with optical-electrical-optical operations. Proposals from \cite{xor3, all-optical-nc} have examined the enabling technologies for performing network coding in photonic domain. Next the works in \cite{hai_comletter, hai_access, hai_oft, hai_comcom, hai_comcom2, hai_springer, hai_rtuwo, hai_systems, hai_springer2, nc_others9, nc_others7} have renewed the interests and demonstrated potential benefits of adding photonic network coding layer to optical networks. The use of digital all-optical physical-layer network coding has also been extended to mm-wave radio-over-fiber networks \cite{nc_others1}. In the realm of visible light communication, there have been increasing works addressing the use of network coding \cite{nc_others2, nc_others3} for boosting capacity and/or transmission qualities. A lately published work \cite{n3} demonstrating channel-aware network coding has achieved around two orders of magnitude reductions in the bit-error-rate (BER). Data-center networking has also been an active area for applying NC to reduce effective traffic \cite{nc_others5, nc_others6}. Recently, physical layer encryption with NC has first been  addressed in \cite{nc_security1, nc_security2} and marked an interesting avenue for re-considering the encryption at scale.  \\
	
	Optical aggregation and de-aggregation have been progressing fast in recent years, permitting two or more lower bit-rate channels being added to a single higher bit-rate \cite{optical_processing_1, optical_processing_2, aggregation1, aggregation2}. The maturing of such optical signal processing technology offers unique opportunities to re-imagine the functions of optical switching node in optical-bypass realm. Specifically, integrating the all-optical aggregation function to optical nodes is expected to reverse the conventional assumption governing optical-bypass networking that each transiting lightpath must be isolated from others to avoid unwanted interference. Indeed, by allowing the optical mixing of transitional optical channels, a new optical networking paradigm, namely, \textit{optical-processing-enabled} networks, has been emerged. To the best of our survey, no previously published works have addressed the concept of \textit{optical-processing-enabled} architecture. Our main contributions are therefore highlighted as followed:
	
	\begin{itemize}
		\item We present an architectural proposal for future optical networks by leveraging the optical mixing operations between transitional lightpaths crossing the same intermediate node.  
		\item We propose the integration of all-optical channel aggregation and de-aggregation operation into optical nodes to attain greater network efficiency 
		\item We develop a mathematical framework for maximizing the spectral benefits in aggregation-aware optical networks 
		\item We showcase numerical results on the realistic network topology COST239 to demonstrate the spectral efficacy of \textit{optical-processing-enabled} networks 
	\end{itemize}

	\section{Optical Channel Aggregation and De-aggregation: Operational Principles and Perspectives}
	In recent years, we have been witnessing a transition in computing moving from electronics to the emerging field of photonics and this has been driven by the massive investment in photonic computing with the promise of solving the Moore's law. Essentially, photonic processing can overcome the fundamental physical limitation of modern electronic processing much in the way that electronic systems surpassing the physical limitations of mechanics. Different from electronic, photonic can make use of all four physical dimensions including amplitude, wavelength, phase and polarization to achieve a wide range of signal processing and computing capabilities. All-optical processing is envisioned therefore as the blueprint for future communication networks to overcome electronic bottleneck and to get in Tb/s regime in a super-efficient way. \\ 
	
	In line with unprecedented advances in photonic processing, all-optical aggregation and de-aggregation technologies have been proposed, investigated and implemented. It has to be noted that the operation of aggregating lower-speed channels into a single higher bit-rate one and de-aggregation constitutes the major task in optical networks and traditionally, this important operation has been performed in the electrical domain. In overcoming the scalability issues of electronic aggregation, various all-optical methods have been proposed based on nonlinear effects. Specifically, the goal of channel aggregation are realized through processes such as four-wave mixing, cross-phase modulation, self-phase modulation and cross-gain modulation which are generated from the second or third-order susceptibility of nonlinear mediums, particularly, highly nonlinear fiber and semiconductor optical amplifiers. There have been increasing experimental works for optical aggregation; for instance, \cite{aggregation4} demonstrated combination of multiple on-off keying (OOK) signals to a MPSK signal, \cite{optical_processing_2} performed adding of two QPSK signals to a 16-QAM signal and \cite{aggregation3} implemented four OOK signals to a polarization division multiplexing (PDM) QPSK signal. \\
	
	A recent experimental works demonstrating remarkable results for 10 Gbaud QAM-16 signal generation from aggregation of two 10 Gbaud QPSK signals has been presented in \cite{aggregation5}. Different from traditional approaches relying on nonlinear effect, this work has marked a departure with a completely linear method for the vector summation of the input channels which is realized by Kerr combs using ring resonators and coherent comb sources like mode-locked lasers. This technique is called coherent vector addition whose merits are notably operational reconfigurability, less system complexity and less power consumption compared to nonlinear-based ones. Figure 1 shows the schematic diagram of adding two QPSK signals into one 16-QAM and decomposing such 16-QAM again into two separate QPSK signals. This operation helps to increase the channel utilization by freeing up the low-speed channels in favorable conditions and thus doubling spectrum efficiency. \\
	
	\begin{figure}[!ht]
		\centering
		\includegraphics[width=\linewidth, height = 7cm]{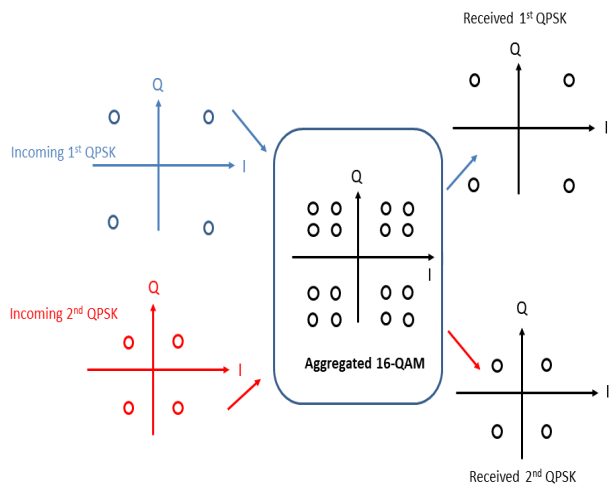}
		\caption{Schematic Illustration of the Aggregation of two QPSK signals into a single 16-QAM one and vice versa}
		\label{fig:i1}
	\end{figure}
	
	\begin{figure}[!ht]
		\centering
		\includegraphics[width=\linewidth, height = 7cm]{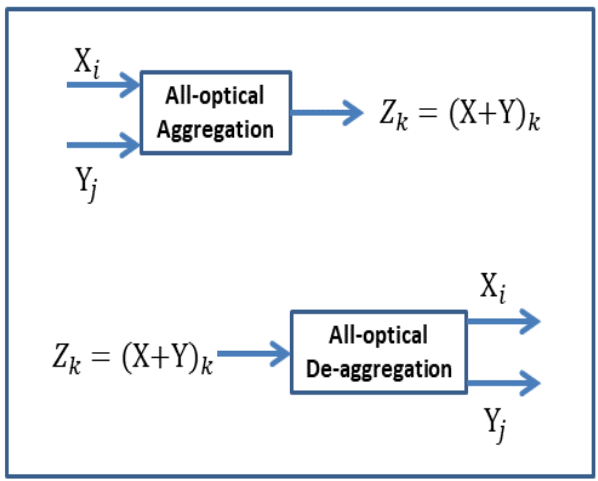}
		\caption{A generic model for optical aggregation and de-aggregation}
		\label{fig:i2}
	\end{figure}
	
	In a forward-looking perspective, accelerated investments and advances in photonic signal processing will be expected to create more possibilities for aggregating optical channels in a programmable manner. Functionally speaking, the aggregation between two (or more) optical channels of different bit-rates and/or formats will be performed in a way that the transmission properties for output signal including wavelength/spectrum and/or modulation format could be subject to selection. As shown in Fig. 2, two optical channels $X$ and $Y$ with transmission parameter $i$ and $j$ (i.e., wavelength/spectrum and modulation format) respectively could be optically aggregated to a single optical channel $Z$ whose line rate is equal to the sum of $X$ and $Y$ and with selectable transmission properties $k$ (e.g., higher-order modulation format). It is expected that the arrival of such optical functionality will have massive impacts to future optical networks including design, planning and management. In the next part, we will highlight the networking impact of incorporating optical aggregation and de-aggregation to optical nodes.

	\section{A New Networking Paradigm: Optical-Processing-Enabled Network}
	The recent major revolution in optical networking architecture has started in the year 2000s with the advent of optical-bypass-enabled networks, permitting elimination of  massive optical-electrical-optical interfaces and therefore, generating substantial cost savings. In optical-bypass networking, the basic functions of an optical node includes add/drop and cross-connect optical channels. Initiatives for improving optical-bypass networking have been proposed by integrating simple optical signal processing functions including regeneration and wavelength/format conversion for each individual transitional lightpath. However, the foundation of optical-bypass networking paradigm remain immutable that each lightpath crossing an intermediate node should be separately from others to mitigate unwanted interference which may degrade the optical signal qualities. In light of tremendous progresses in optical signal processing permitting precisely controlled interference between optical channels, superposition of transiting lightpaths traversing the same optical node pave the way for redefining the optical network architecture. We highlight this perspective by leveraging the integration of all-optical aggregation and de-aggregation capability through the following example. \\
	
	Assuming that there are two traffic demands $a$ and $b$ with 100 Gbps from node $A$ to node $C$ and from node $B$ to node $C$ respectively. In optical-bypass networking, accommodating such two requests involves finding the route and assigning the appropriate wavelength for each demand and it is shown in Fig. 3. It is noted that the two lightpaths carrying traffic for demand $a$ and $b$ are routed over the same link $XI$ and $IC$, and more importantly, both are optically crossed node $X$. In terms of wavelength count across link $XI$ and $IC$ two wavelengths are thus needed. \\
	
	\begin{figure}[!ht]
		\centering
		\includegraphics[width=0.8\linewidth, height = 6cm]{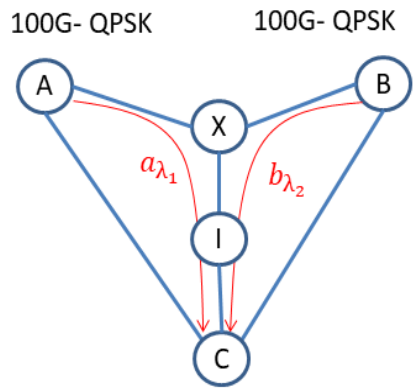}
		\caption{Traffic Provisioning in Optical-bypass Networking}
		\label{fig:i3}
	\end{figure}
	
	We turn the attention to the \textit{optical-processing-enabled} paradigm where all-optical aggregation at node $X$ is enabled. In this paradigm, two transitional lightpaths $a_{\lambda_1}$ and $b_{\lambda_2}$ could be optically mixed at node $X$. Specifically, two 100G QPSK signals modulated on wavelength $\lambda_1$ and $\lambda_2$ respectively are optically aggregated to a single higher-order modulation format signal, that is, 200G 16-QAM, modulated on wavelength $\lambda_3$. By aggregating lower-capacity channels into a single higher-capacity channel of higher-order modulation format, greater spectrum efficiency could be realized. At the common destination node $C$, the aggregated signal could be optically decomposed into individual ones as shown in Fig. 4. \\
	
	\begin{figure}[!ht]
		\centering
		\includegraphics[width=\linewidth, height = 6.5cm]{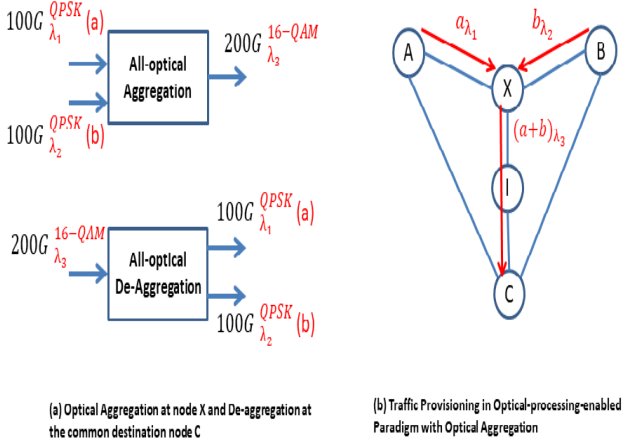}
		\caption{Optical-processing-enabled Paradigm with Optical Aggregation and De-aggregation}
		\label{fig:i4}
	\end{figure}
	
	It should be noted that the \textit{optical-processing-enabled} paradigm introduces more networking flexibility by permitting the controlled interference among two or more favorable lightpaths and this poses important ramifications for network design algorithms to maximize the potential benefits. For the case of optical aggregation above, the critical problem to be addressed is the determination of which pair of demands could be optically aggregated, where are the nodes for optical aggregation and what are the routes for aggregated lightpaths. In the next section, we provide an optimal mathematical framework for solving such problem. \\

	\section{A Mathematical Formulation for Optimal Aggregation-aware Routing}
	In order to leverage the use of optical aggregation and de-aggregation in network design and planning, we restrict our studies to the practical and simple setting where the aggregation is only performed between two demands of the same line-rate and modulation format while the de-aggregation takes place only at the common destination node. This can be achieved by making use of a pretty maturing optical aggregator combing two QPSK signals to a single 16-QAM channel \cite{aggregation5, optical_processing_2} \\
	
	The spectral saving enabled by optical aggregation and de-aggregation is reliant on the underlying physical topology, the nature of traffic demands, routing schemes together with the location of aggregation nodes. In this section, we develop a mathematical formulation that could minimize the wavelength link cost or alternatively maximize the impact of aggregation. Inputs to the model are the physical topology, traffic demands and the output is the optimal identification of routing for each demand, set of pair of demand for aggregation and their respective aggregation nodes so that the wavelength link cost is minimized. \\

	\noindent{Inputs:}
	\begin{itemize}
		\item $G(V,E)$: A fiber-optic network composes $|V|$ nodes and $|E|$ optical fiber links. Each link $e \in E$ is identified by its beginning node $s(e)$ and its ending node $r(e)$. 
		\item $D$: Set of traffic demands; each demand $d \in D$ consists of its source node $s(d)$ and destination node $r(d)$ respectively. All demands $d$ request equally an amount of traffic equivalent to a wavelength capacity (e.g., 100 Gbps)
	\end{itemize}
	
	\noindent{Variables:}
	\begin{itemize}
		\item $x_{e}^{d}  \in \{0,1\} $: equals 1 if link $e$ is used for routing traffic of demand $d$, 0 otherwise.\\
		\item $z_{e}^{d, v} \in \{0,1\} $: equals 1 if demand $d$ is aggregated at node $v$ and link $e$ belongs to the routing path of the aggregated lightpath, 0 otherwise \\
		\item $\theta_{v}^{d} \in \{0, 1\} $: equals 1 if demand $d$ is aggregated at node $v$, 0 otherwise \\
		\item $f_{d_1}^{d_2} \in \{0, 1\} $: equals 1 if demand $d_1$ is aggregated with demand $d_2$, 0 otherwise\\
		
	\end{itemize}
	
	\noindent{Objective Function: Minimize the following equation}
	
	\begin{equation} \label{eq:obj1}
		\sum_{d \in D} \sum_{e \in E} x_{e}^{d} - \sum_{d \in D} \sum_{e \in E} \sum_{v \in V}  \frac{z_{e}^{d, v}}{2} \\
	\end{equation}

	\noindent{subject to the following constraints:}
	
	\begin{equation} \label{eq:c3} 
		\begin{split}
			\sum_{e \in {E}: v \equiv s(e)} x_{e}^{d} -\sum_{e \in {E}: v \equiv r(e)} x_{e}^{d}= \\		
			\begin{cases} 
				1 &\mbox{if } v \equiv s(d) \\ 
				-1 & \mbox{if } v \equiv r(d)\\
				$0$ & otherwise \\
			\end{cases}     \qquad \qquad \forall d \in D, \forall v \in V \hfill
		\end{split}
	\end{equation}
	
	\begin{align} \label{eq:c6}  {
			\sum_{v \in V} \theta_{v}^{d} \leq 1 \qquad and \qquad \theta_{v}^{d} = 0 \qquad \mbox{if } v \equiv r(d) \qquad \forall  d \in D
		}
	\end{align}
	
	\begin{align} \label{eq:c7} {
			\sum_{d_2 \in D} f_{d_1}^{d_2} \leq 1 \qquad \forall d_1 \in D
		}
	\end{align}
	
	\begin{equation} \label{eq:c8}
		\sum_{d_2 \in D: r(d_2) \neq r(d_1)}  {f^{d_1}_{d_2}} = 0 \qquad \forall d_1 \in D
	\end{equation}

	\begin{align} \label{eq:c9} {
			f_{d_1}^{d_2} = f_{d_2}^{d_1} \qquad \forall d_1, d_2 \in D
		}
	\end{align}
	
	\begin{align} \label{eq:c10} {
			\sum_{v \in V} z_{e}^{d_1, v} \leq \sum_{d_2 \in D} f_{d_1}^{d_2} \qquad \forall d_1 \in D, \forall e \in E
		}
	\end{align}
	
	\begin{align} \label{eq:c11} {
			\sum_{d_2 \in D} f_{d_1}^{d_2} = \sum_{v \in V} \theta_{v}^{d_1} \qquad \forall d_1 \in D
		}
	\end{align}

	\begin{align} \label{eq:c14} {
			\theta_{v}^{d_1} - \theta_{v}^{d_2}+f_{d_1}^{d_2} \leq 1 \qquad \forall d_1, d_2 \in D, \forall v \in V
		}
	\end{align}
	
	\begin{align} \label{eq:c15} {
			\theta_{v}^{d_2} - \theta_{v}^{d_1}+f_{d_1}^{d_2} \leq 1 \qquad \forall d_1, d_2 \in D, \forall v \in V
		}
	\end{align}
	
	\begin{align} \label{eq:c16} {
			z_{e}^{d, v}  \leq x_{e}^{d}  \qquad \forall d \in D, \forall v \in V, \forall e \in E
		}
	\end{align}
	

\begin{equation} \label{eq:c17}
	\begin{split}
		\sum_{e \in E: i \equiv s(e)} z_{e}^{d, v} - \sum_{e \in E: i \equiv r(e)} z_{e}^{d, v}= \\
		\begin{cases} 
			\theta_{v}^{d} &\mbox{if } i \equiv v \\ 
			-\theta_{v}^{d} & \mbox{if } i \equiv r(d)\\
			0 & \mbox{otherwise} \\
		\end{cases} \qquad \qquad \forall d \in D, \forall v \in V, \forall i \in V \hfill
	\end{split}
\end{equation}

The objective in Eq. \ref{eq:obj1} aims to minimize the wavelength link cost for accommodating all demands. Note that whenever there is an aggregation operation, it results in a saving of wavelength links. Equation \ref{eq:c3} is the conventional flow conservation constraint for each demand. Constraint in Eq. \ref{eq:c6} captures the rule that each demand could be aggregated at most one times at a certain aggregation node and that node has to be different from the destination. The condition that each demand is aggregated with no more than one another demand having the same destination is guaranteed by Eqs. \ref{eq:c7}, \ref{eq:c8}, and \ref{eq:c9}. Constraints in Eqs. \ref{eq:c10}, \ref{eq:c11} are for coherence purpose connecting each aggregated demand with the respective aggregation node and aggregation links. Constraints (\ref{eq:c14}, \ref{eq:c15}) ensures the observation that if two demands are aggregated, the aggregation operation has to be occurred at the same node. Constraint in Eq. \ref{eq:c16} specifies the routing for aggregated lightpaths. Final constraint (\ref{eq:c17}) is for conservation of aggregated flows.

\section{Numerical Results}
\begin{figure}[!ht]
	\centering
	\includegraphics[width=0.8\linewidth, height = 7cm]{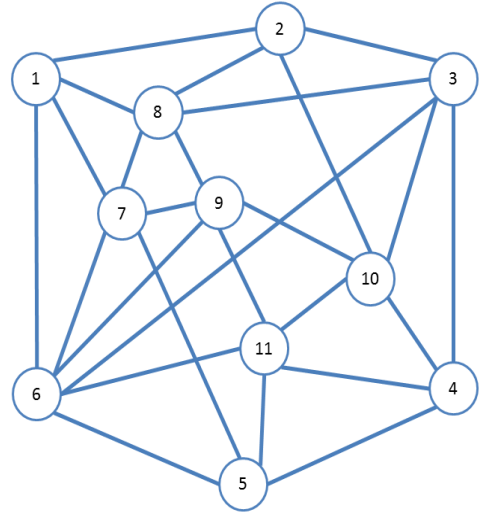}
	\caption{Network Topology under Test}
	\label{fig:cost239}
\end{figure}

This section presents numerical evaluations comparing our proposal that leverages the use of optical aggregation within the framework of \textit{optical-processing-enabled} networks to the traditional optical-bypass networking. The comparison is drawn on a realistic COST239 network topology shown in Fig. \ref{fig:cost239}. In order to exploit aggregation opportunities among demands, the traffic under consideration is generated following the two-to-many model where two nodes are randomly selected as sources and many remaining nodes are also arbitrarily chosen as the destinations. For this traffic scenario, favorable conditions for aggregation operation have been created as there have been multiple pair of demands sharing the same destination. We consider three increasing loads corresponding to 5, 7 and 9 destination nodes and for each load, 10 traffic samples are simulated. Each traffic request is served by allocating a dedicated wavelength channel capacity in the standard C-band where channel spacing is 50 $GHz$. The performance metric for comparison is the wavelength link cost for accommodating all demands. 

Figure 6, 7, and 8 report the performance comparison between our proposal of aggregation-aware routing in optical-processing-enabled networks and the traditional routing in optical-bypass networks for three traffic loads. As a general trend to observe, the aggregation-aware scheme is highly more spectrum-efficient than the non-aggregation counterpart. In all simulated scenarios, the maximum relative gain is up to slightly more than $30\%$. \\

\begin{figure}[!ht]
	\centering
	\includegraphics[width=\linewidth, height = 6cm]{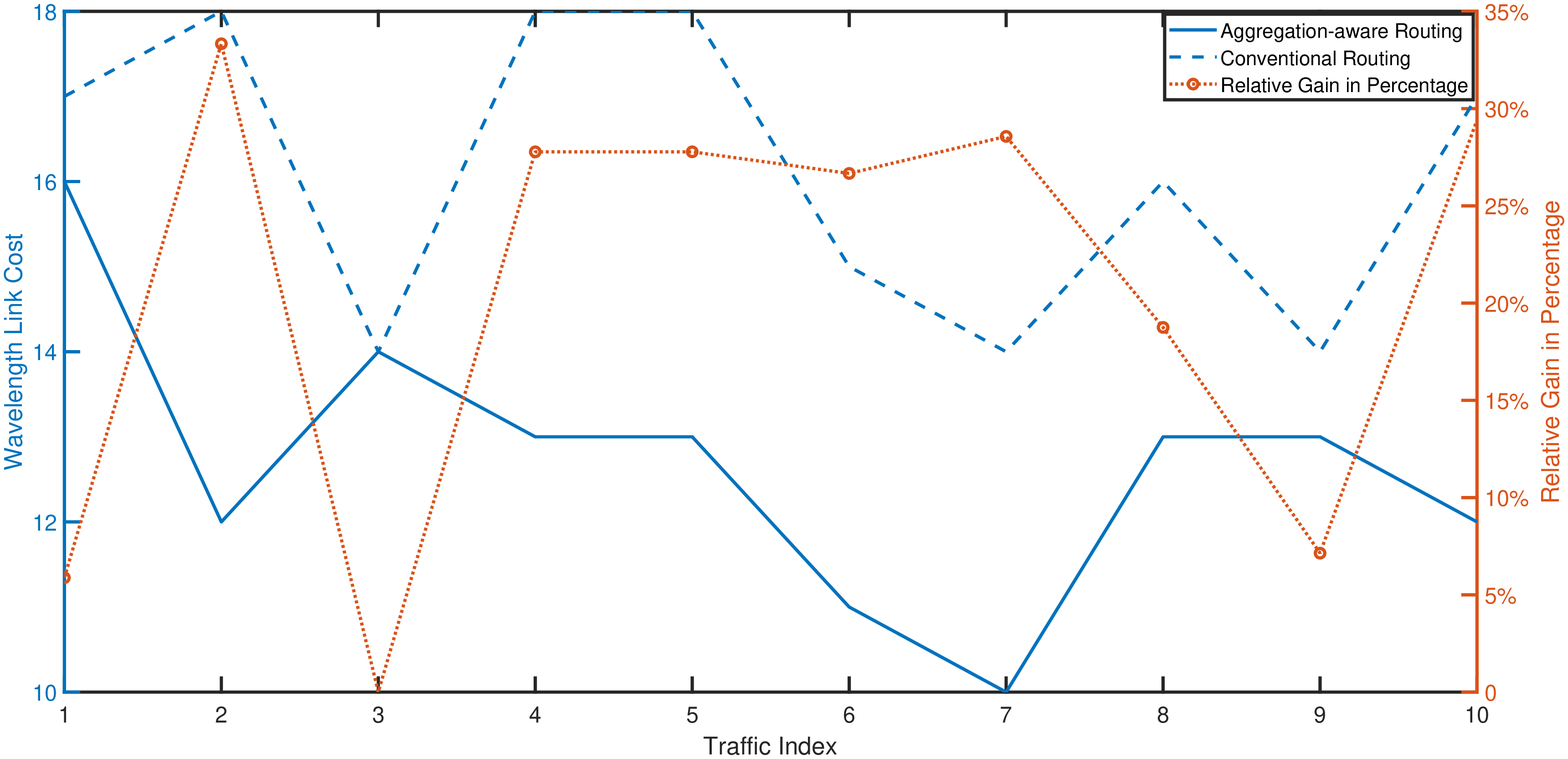}
	\caption{Routing Cost Comparison for Low-load Traffic}
	\label{fig:i5}
\end{figure}

\begin{figure}[!ht]
	\centering
	\includegraphics[width=\linewidth, height = 6cm]{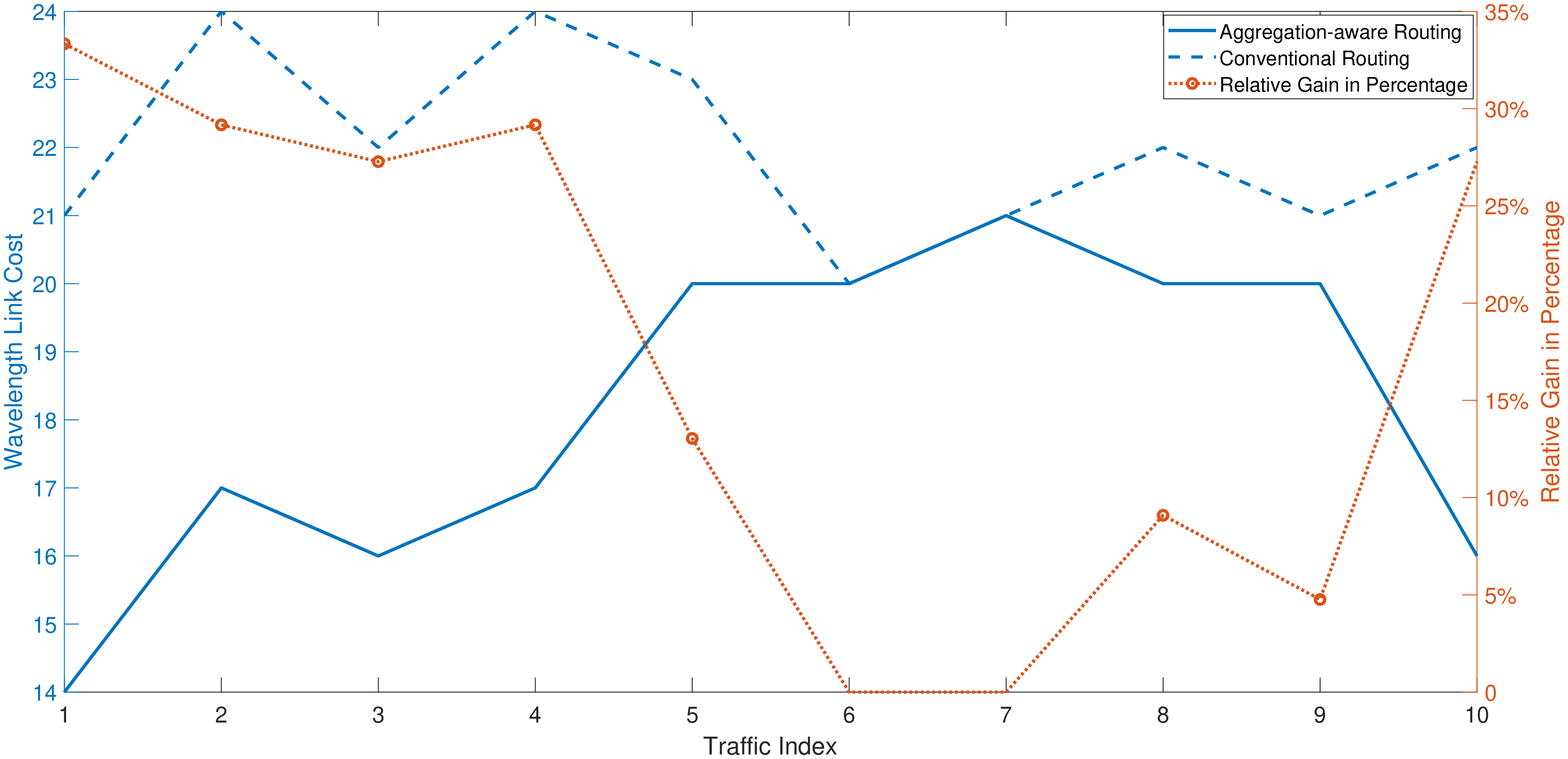}
	\caption{Routing Cost Comparison for Medium-load Traffic}
	\label{fig:i6}
\end{figure}

An interesting observation is that there are cases that aggregation-aware routing does not result in spectral gain compared with conventional routing. This can be observed at traffic index 3 for the low-load scenario in Fig. 6, traffic index 6 and 7 for the medium-load cases in Fig. 7. It might be explained that in order to perform optical aggregation between two demands, in addition to the necessary condition of having the same destination, two demands must share a segment of its route. Such constraint might cause their routing diverged from the shortest ones and if the aggregation benefit could not compensate for the divergence from the shortest routes, no gain will be realized. \\

\begin{figure}[!ht]
	\centering
	\includegraphics[width=\linewidth, height = 6cm]{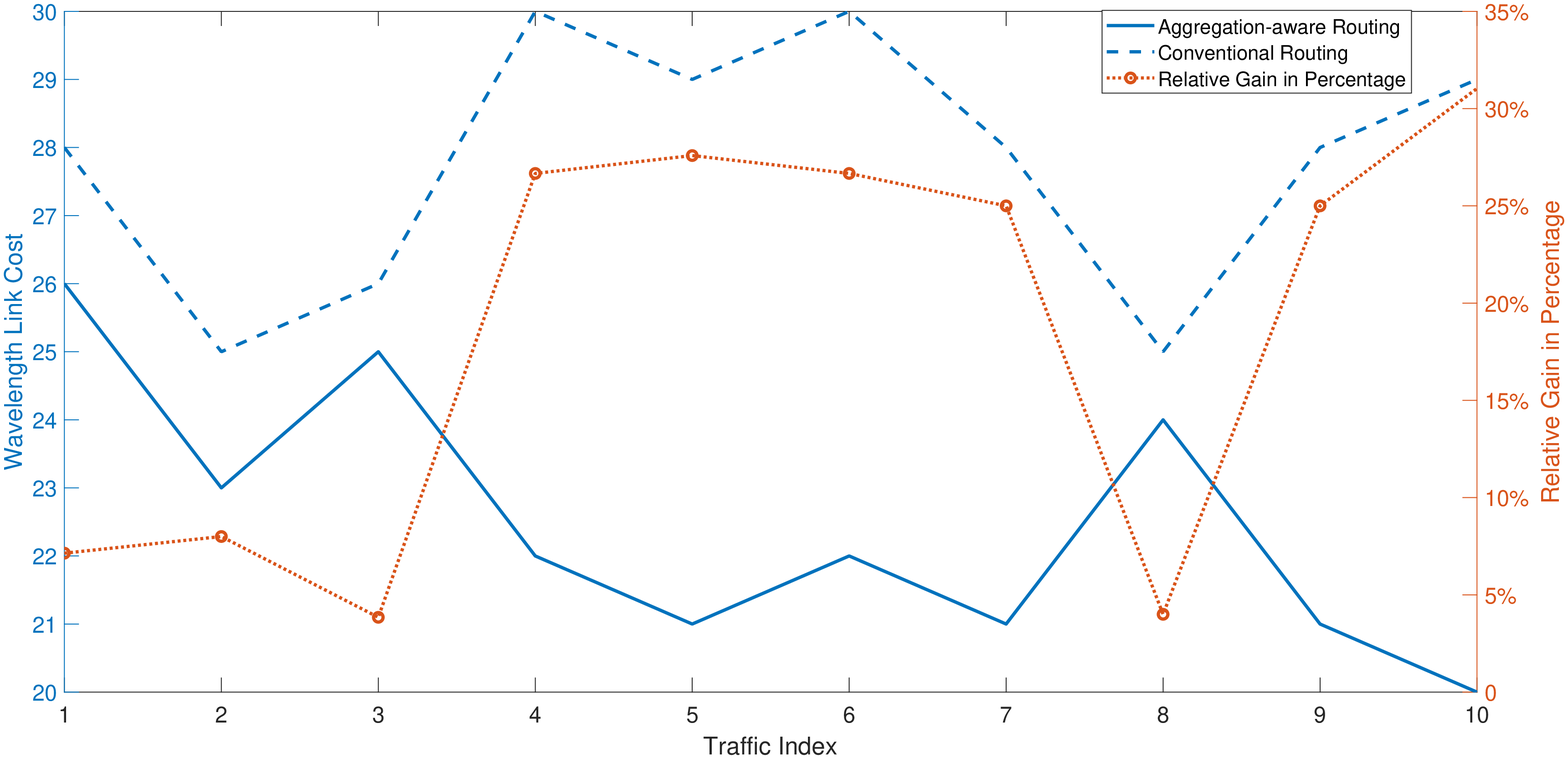}
	\caption{Routing Cost Comparison for High-load Traffic}
	\label{fig:i7}
\end{figure}

For the high-load traffic in Fig. 8, aggregation-aware mechanism is strictly better than the traditional routing. Nevertheless, the average gain across three traffic loads are quite comparable, being $20.5\%$, $17.3\%$ and $18.5\%$ for low-load, medium-load and high-load respectively. \\

To inspect more on the mechanism of aggregation-aware routing, Table 1 and 2 present the comprehensive simulation results for one traffic sample of low-load scenario. It is noted that for aggregation-aware routing, in addition to the routing information, the determination of aggregation node, pair of demands for aggregation, and the links for routing aggregated lightpaths are also revealed. Here we would like to draw the attention to the routing of demand $9\rightarrow2$ and $9\rightarrow4$ as it uncovers an important observation. As shown in Tab. 2, both two demands select the shortest path composing of two links to minimize the wavelength link cost. However, in aggregation-aware mechanism shown in Tab. 1, these two demands are routed over longer paths consisting of three links to make them aggregated to other demands $7\rightarrow2$ and $7\rightarrow4$. This is justifiable because the spectral saving enabled by aggregation is higher than the cost of deviation from shortest paths.   

\begin{table*}[ht]
	\caption{Aggregation-aware Routing}
	\label{tab: r1}
	\centering
	\begin{tabular}{|c|c|c|c|c|}
		\hline
		Demand & Routing & Aggregation Node & Aggregation Links & With Demand \\
		\hline 
		7$\rightarrow$2 & (7-1-2) & 7 & (7-1-2) & 9$\rightarrow$2 \\
		7$\rightarrow$11 & (7-9-11) & 9 & (9-11) & 9$\rightarrow$11 \\
		7$\rightarrow$4 & (7-5-4) & 7 & (7-5-4) & 9$\rightarrow$4 \\
		7$\rightarrow$3 & (7-8-3) & 8 & (8-3) & 9$\rightarrow$3 \\
		7$\rightarrow$1 & (7-1) & 7 & (7-1) & 9$\rightarrow$1 \\
		
		9$\rightarrow$2 & (9-7-1-2) & 7 & (7-1-2) & 7$\rightarrow$2 \\
		9$\rightarrow$11 & (9-11) & 9 & (9-11) & 7$\rightarrow$11 \\ 
		9$\rightarrow$4 & (9-7-5-4) & 7 & (7-5-4) & 7$\rightarrow$4 \\
		9$\rightarrow$3 & (9-8-3) & 8 & (8-3) & 7$\rightarrow$3 \\
		9$\rightarrow$1 & (9-7-1) & 7 & (7-1) & 7$\rightarrow$1 \\
		
		\hline
	\end{tabular}
\end{table*}

\begin{table*}[ht]
	\caption{Conventional Routing in Optical-bypass Networks}
	\label{tab: r2}
	\centering
	\begin{tabular}{|c|c|}
		\hline
		Demand & Routing \\
		\hline 
		7$\rightarrow$2 & (7-1-2)  \\
		7$\rightarrow$11 & (7-9-11)  \\
		7$\rightarrow$4 & (7-5-4)  \\
		7$\rightarrow$3 & (7-8-3)  \\
		7$\rightarrow$1 & (7-1)  \\
		
		9$\rightarrow$2 & (9-8-2)  \\
		9$\rightarrow$11 & (9-11)  \\ 
		9$\rightarrow$4 & (9-10-4)  \\
		9$\rightarrow$3 & (9-8-3)  \\
		9$\rightarrow$1 & (9-8-1)  \\
		
		\hline
	\end{tabular}
\end{table*}


\section{Conclusions}
The aim of this paper was to propose a new architectural paradigm for future optical networks that leverage the optical signal processing between transiting lightpaths to attain greater capacity efficiency. As a departure from simple functions of add/drop and cross-connect in optical-bypass networking, our proposal\textemdash optical-processing-enabled networks\textemdash took advantage of optical aggregation capability that permits adding lower-speed channels into a single higher bit-rate one for capacity savings. Specifically, we made use of an optical aggregator that can combine two QPSK channels into a single 16-QAM one and developed an optimal mathematical model based on integer linear programming to maximize the aggregation benefits. Numerical results on a realistic topology were provided to verify the efficacy of our proposal compared to the conventional approach and it was shown that up to more than $30\%$ spectral saving could be attained in favorable conditions. \\

Albeit still at early-stage, this paper has identified an important observation that by permitting optical superposition of transitional lightpaths, new opportunities for capacity saving could be emerged thanks to reducing the effective traffic load in the network. In fact, optical channel aggregation and de-aggregation has begun to alter the way we think about optical networks architecture, reversing the conventional assumption that transitional lightpaths should be maximally separated to avoid unwanted interferences. Driven by the rapid advances in photonic signal processing technologies and the great incentive from operators in making the optical networks more spectrum-efficient and cost-efficient, further innovations in optical-processing-enabled realm from both enabling technologies to designing algorithms should be pushed to fully unlock its practicability and applicability.      

\label{sect: conclusion}

\section*{Conflict of interest}
The authors declare that they have no conflict of interest.

\section*{Data availability}
The datasets generated during and/or analysed during the current study are available from the corresponding author on reasonable request. 

\bibliographystyle{spmpsci_modified}      
\bibliography{ref}   

\end{document}